\documentclass[]{spie}  %>>> use for US letter paper
\usepackage{amsmath,amsfonts,amssymb}
\usepackage{graphicx}
\usepackage[colorlinks=true, allcolors=blue]{hyperref}
\usepackage{comment}
\usepackage{subcaption}

\title{Design and Testing of the Motorized 2-DoF Folding Mirror 1 for the VLT BlueMUSE Instrument}

\author[a]{Gloria Mellinand}
\author[a]{Diane Chapuis Kerouanton}
\author[a]{Malak Galal}
\author[a]{Axel Nicolier}
\author[a]{Aurélien Genin}
\author[a]{Zeno Amann}
\author[a]{Sébastien Pernecker}
\author[b]{Rémi Giroud}
\author[b]{Alexandre Jeanneau}
\author[b]{Florence Laurent}
\author[b]{Johan Richard}
\author[a]{Jean-Paul Kneib}
\affil[a]{Institute of Physics, Laboratory of Astrophysics, Ecole Polytechnique Fédérale de Lausanne (EPFL), Observatoire de Sauvergny, CH-1290 Versoix, Switzerland}
\affil[b]{Univ Lyon, Univ Lyon1, Ens de Lyon, CNRS, Centre de Recherche Astrophysique de Lyon UMR5574, F-69230,
Saint-Genis-Laval, France}

\authorinfo{Further author information: (Send correspondence to G. M. and D.C.)\\ G.M.: E-mail: gloria.mellinand@epfl.ch, Telephone: +41 21 693 26 71\\ D.C.: diane.chapuis@epfl.ch, Telephone: +41 21 693 54 37}

\begin{document} 
\maketitle

\begin{abstract}
BlueMUSE is a blue-optimized, medium spectral resolution, panoramic integral field spectrograph under development for the Very Large Telescope (VLT). The project is now fully entering the design phase. With an optimized transmission down to 350 nm, spectral resolution of R$\sim$3500 on average across the wavelength range, and a large FoV (1 arcmin²), BlueMUSE will open up a new range of galactic and extragalactic science cases facilitated by its specific capabilities.

In this paper, we present the design, implementation and evaluation of the motorized mount developed for Folding Mirror 1 (FM1) of the VLT BlueMUSE instrument. The mount provides two degrees of freedom in tip and tilt and is engineered to correct misalignments caused by environmental variations. The preliminary results demonstrate that the motorized FM1 tilt mount achieves the required precision and maintains alignment stability within the tight tolerances defined by the BlueMUSE project.
\end{abstract}


\section{INTRODUCTION}
\label{sec:intro}  % \label{} allows reference to this section

\begin{figure}[!h]
    \centering
    \includegraphics[scale=0.4]{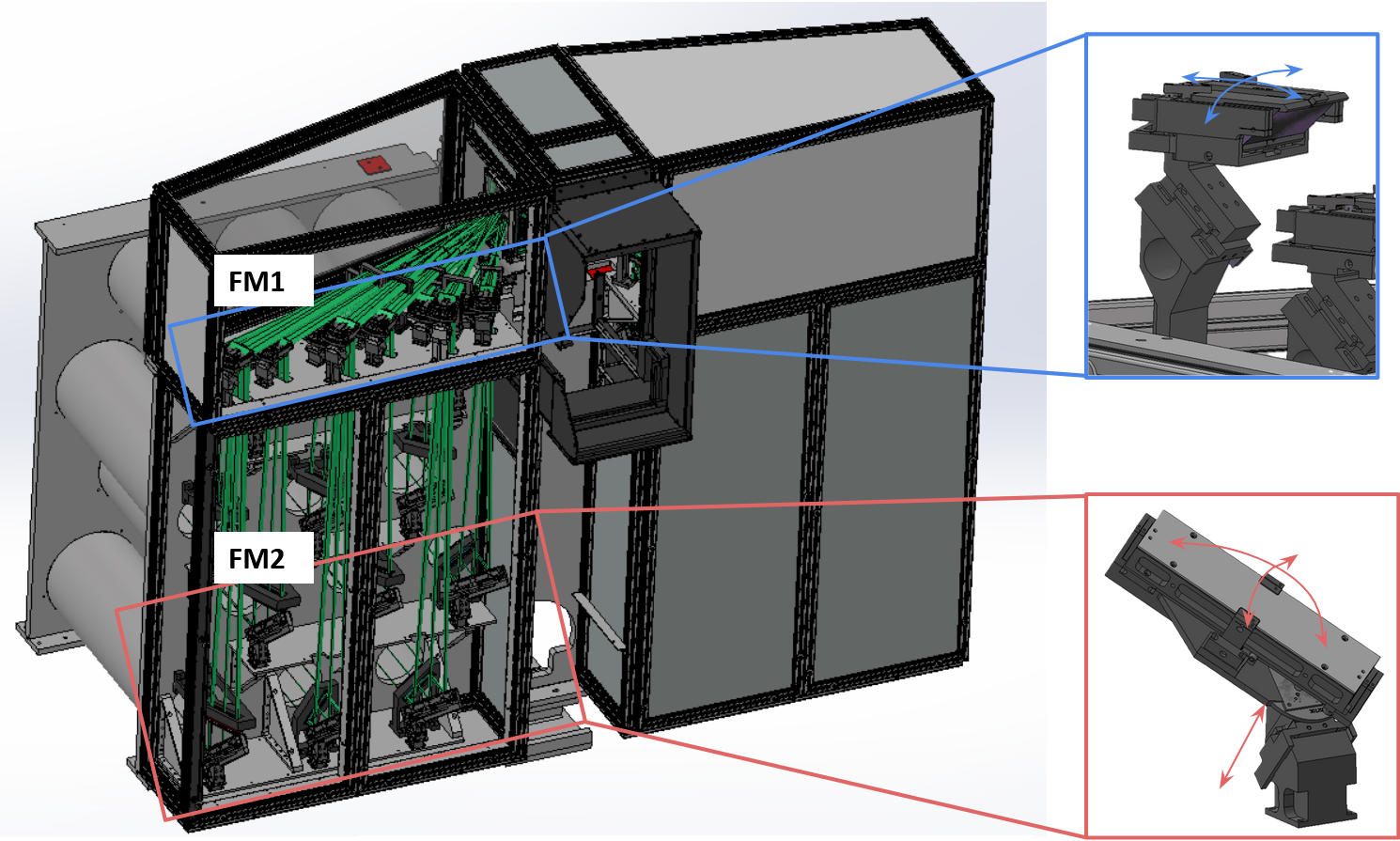}
    \caption{CAD view of the MUSE instrument and the Folding Mirrors (FM) within. The optical paths are shown in green, and the left hand side of the instrument is mirrored on the right. Blue arrows show the FM1's rotational degrees of freedom, while red arrows show the FM2's addition of a translational degree of freedom}  % Single overall caption
    \label{fig:MUSE_FM}
\end{figure}

BlueMUSE, building on the success of its predecessor MUSE \cite{Bacon2010MUSE}\cite{Bacon2015MUSE}, aims to perform high-resolution spectroscopic imaging of the cosmos using a set of custom blue-optimized spectrographs and a complex relay optics architecture \cite{Richard2024BlueMUSE}\cite{jeanneau2020curved}. Once installed on the Very Large Telescope (VLT) and achieving first light in 2034, BlueMUSE will focus on observations in the blue portion of the visible spectrum, enabling scientific studies ranging from young stellar populations to faint diffuse emissions in the universe\cite{Richard2021BlueMUSE}. While targeting shorter wavelengths, BlueMUSE benefits from a strong architectural heritage from MUSE, as well as from the operational experience and lessons learned from the previous instrument generation. 

A critical contributor to the optical performance of the instrument is the thermal stability of the 32 Folding Mirrors (FMs) distributed throughout the relay optics system \cite{Cai2024MUSEFEA}. Two types of folding mirrors are implemented in the instrument. Folding Mirror 1 (FM1) provides two rotational degrees of freedom corresponding to $\Theta Y_{FM}$ (tip) and $\Theta X_{FM}$ (tilt) motions of the mirror surface. Folding Mirror 2 (FM2) includes the same rotational degrees of freedom, with the addition of a translational adjustment perpendicular to the mirror surface\cite{Mellinand2026FM2}. The instrument contains a total of 16 FM1 units and 16 FM2 units. 

Although the VLT site at Cerro Paranal benefits from a relatively stable arid climate with limited annual thermal variations\cite{sarazin1999climate}, even small temperature changes between daytime and nighttime can induce significant thermo-mechanical deformations in precision optical mounts \cite{fischer2008optical}\cite{NOETHE20021}. These deformations can distort the optical beam path and require frequent recalibration of the instrument \cite{Weilbacher2020data}. The MUSE instrument currently employs manually adjusted folding mirror mounts, as seen in Figure \ref{fig:MUSE_FM}, for which thermal variations at dusk can generate several arcsecond-level beam deviations. Correcting these misalignments requires manual recalibration by trained operators, a time-consuming and delicate procedure that needs to be performed at night between exposures. 

These limitations motivate the development of a compact, automated and thermally stable motorized platform for folding mirrors. Such a system would enable rapid mirror repositioning while maintaining optical stability under realistic environmental variations, thereby reducing the need for frequent manual intervention. Achieving arcsecond-level repeatability and long-term stability is a significant engineering challenge, as numerous factors contribute to positioning errors, including mechanical backlash, material thermal expansion, assembly tolerances, and structural compliance \cite{giesen2003design}. These effects must be mitigated through careful mechanical design, material selection, motion reduction strategies, and control architecture optimization. Furthermore, validating such performance requires a highly accurate test bench capable of characterizing angular stability at the arcsecond scale. 

This paper presents the system architecture and mechanical design of the proposed FM1 motorized mount, with particular emphasis on its two rotational axes. We then describe the first stainless-steel prototype and the dedicated experimental setup developed to evaluate the system according to repeatability and short and long-term thermal stability requirements. Finally, we discuss the experimental results obtained with the prototype and outline future developments for the next iteration of the Folding Mirror 1 design.

\newpage
\section{Folding Mirror 1 motorized mount hardware architecture}
\begin{figure}[!h]
    \centering
    \includegraphics[width=0.5\textwidth]{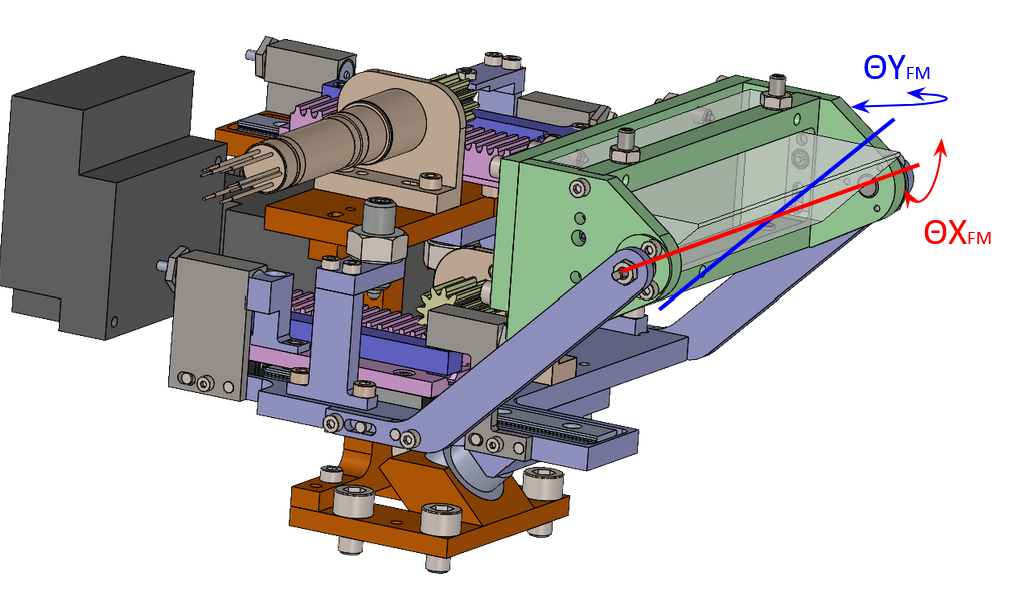}
    \caption{Folding mirror 1 (FM1) opto-mechanical assembly. The $\Theta X_{FM}$ and $\Theta Y_{FM}$ axes of rotation are displayed on the mirror surface.} 
    \label{fig:FM1_new}
\end{figure}

The FM1 mechanism is actuable in $\Theta X_{FM}$ and $\Theta Y_{FM}$ rotations. The axes of rotation are shown in Figure \ref{fig:FM1_new} and are aligned along the mirror surface.  

The desired performance defining the Folding Mirrors' design and performance describe repeatability, accuracy and range of motion. The following table shows preliminary identified requirements for the FM1 motorized mount:

\begin{table} [!h]
\centering
\begin{tabular}{lcc}
\hline\hline
Requirement & FM1 $\Theta Y_{FM}$ & FM1 $\Theta X_{FM}$ \\
\hline
Repeatability in one step & 2.5 arcsec & 2.5 arcsec \\
Mount stability during operation & 0.5 arcsec & 0.5 arcsec   \\
Long-term mount stability & 0.5 arcsec & 0.5 arcsec   \\
Range & 5 arcmin & 5 arcmin \\
\hline\end{tabular}
    \caption{FM1 motorized mount performance requirements updated in January 2026}
    \label{tab:realignment_req}
\end{table}
In the previous phase, as a first step, off-the-shelf goniometers were tested, but none fit the requirements. Therefore, we came to the conclusion that each mount requires a custom development tested with a prototype, together with the necessary electronics and control software. This need for a custom design is emphasized with the recent update of the requirement for the mount stability during operations, under a maximum temperature variation of ± 2°C, going from 5 arcsec to only 0,5 arcsec.

The current design is displayed in Figure~\ref{fig:FM1_new}

\begin{figure}[!h]
    \centering
    \begin{subfigure}[b]{0.45\textwidth}
        \centering
        \includegraphics[width=\textwidth]{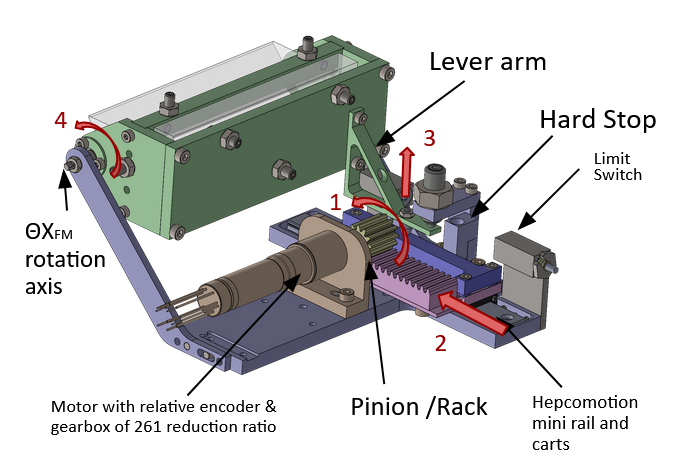}
        \caption{}  
        \label{fig:Theta_X}
    \end{subfigure}
    \hfill
    \begin{subfigure}[b]{0.45\textwidth}
        \centering
        \includegraphics[width=\textwidth]{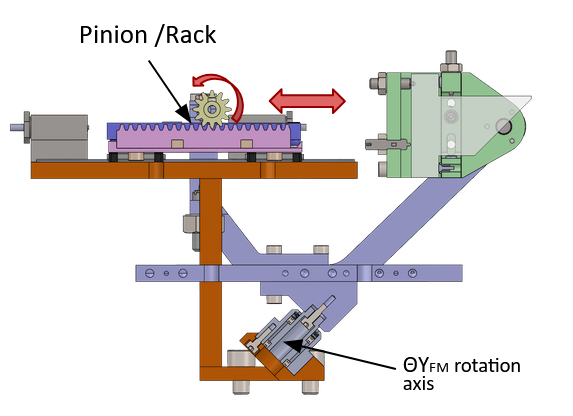}
        \caption{}  
        \label{fig:Theta_Y}
    \end{subfigure}
    \caption{Folding mirror 1 (FM1) mechanical principle: (a) FM1 $\Theta X_{FM}$ stage with red arrows illustrating the motions of components; (b) Cross section of FM1 $\Theta Y_{FM}$ rotation axis}
    \label{fig:rotation_mechanisms}
\end{figure}

\subsection{Mechanical Architecture and Design}
\label{sec:mechanical}
\begin{figure}[!h]
    \centering

    \begin{subfigure}{0.9\textwidth}
        \centering
        \includegraphics[width=\textwidth]{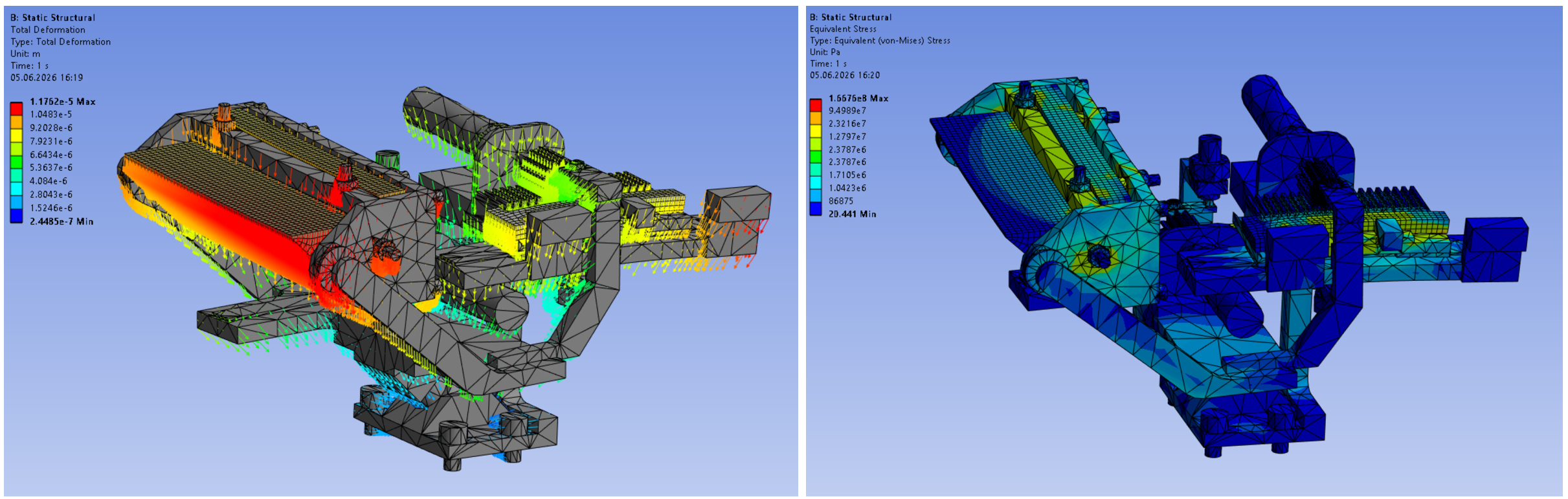}
        \caption{}
        \label{fig:FEA_deformation}
    \end{subfigure}

    \vspace{0.5cm}

    \begin{subfigure}{0.9\textwidth}
        \centering
        \includegraphics[width=\textwidth]{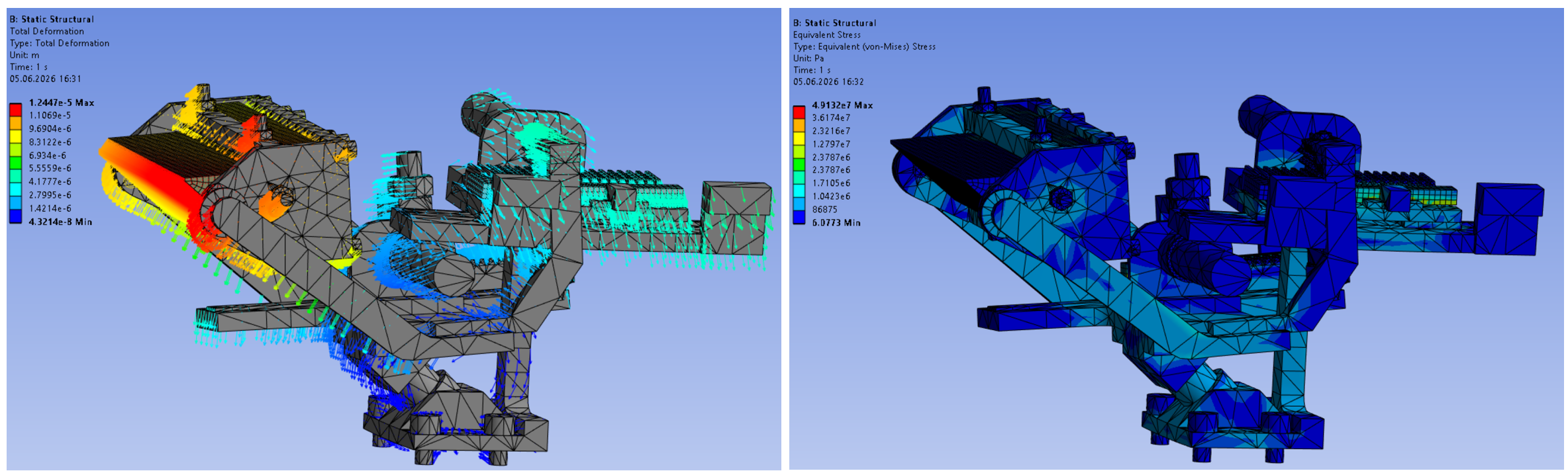}
        \caption{}
        \label{fig:FEA_stress}
    \end{subfigure}

    \caption{Folding mirror 1 (FM1): (a) FM1 displacement and stress FEA with mirror mount in stainless steel; (b) FM1 displacement and stress FEA with mirror mount in Invar 36.}
    \label{fig:FEA}
\end{figure}
To fulfill the requirements listed in Table~\ref{tab:realignment_req}, a large motion reduction is required between the motor rotation and the resulting angular displacement of the mirror. This reduction is necessary to limit the influence of motor resolution, gearbox backlash, manufacturing tolerances, and mechanical play on the final mirror position.

The FM1 mechanism provides two rotational degrees of freedom, $\Theta X_{FM}$ and $\Theta Y_{FM}$, corresponding to tip and tilt motions of the mirror surface. The two rotation axes are aligned with the mirror surface and are implemented through two compact actuation stages stacked within the available mechanical envelope.

For each axis, the motor rotation is first converted into a linear displacement using a rack-and-pinion mechanism, as shown in Figure~\ref{fig:Theta_X}. The rack is attached to a linear carriage carrying a high-stiffness EKasic ceramic slope. The slope inclination is created by a fine shim, producing a height variation of $0.15~\mathrm{mm}$ over a length of $64~\mathrm{mm}$. A ball transfer unit remains in contact with this inclined surface and transmits the vertical displacement to a lever arm directly connected to the mirror support. The resulting lever motion produces the angular rotation of the mirror.

The contact between the ball transfer unit and the slope is maintained by a preload spring. This preload ensures continuous contact during operation and contributes to the robustness of the mechanism under external vibrations, including accelerations associated with earthquake load cases. The mirror itself is constrained through spring-loaded contacts, providing a stable mounting while reducing the risk of parasitic vibrations or over-constraint at the mirror interface.

The mechanism provides a theoretical transmission ratio of

\[
\eta = \frac{\theta_{mirror}}{\theta_{motor}} = 0.153~\mathrm{arcsec/rad},
\]

where $\theta_{motor}$ is the motor rotation before the reducer and $\theta_{mirror}$ is the resulting mirror rotation. This high reduction ratio enables very fine angular positioning. The current geometry provides a theoretical angular range of approximately $6~\mathrm{arcmin}$, compatible with the required $5~\mathrm{arcmin}$ range. The slope angle can be adjusted by changing the shim thickness, allowing the motion range and reduction ratio to be tuned if required.

The same actuation principle is used for the $\Theta Y_{FM}$ axis as seen in Figure \ref{fig:Theta_Y}. Both the $\Theta X_{FM}$ and $\Theta Y_{FM}$ stage use a motor, a rack-and-pinion conversion, a linear carriage, a slope, a ball-bearing contact, and a lever arm connected to the mirror support. The main difference between the two axes lies in the location of the rotation axis within the stacked mechanical architecture. The $\Theta Y_{FM}$ axis uses a set of bearings located near the bottom of the structure, aligned with the mirror surface, while the $\Theta X_{FM}$ axis is integrated in the upper stage.

A first thermo-mechanical finite element analysis (FEA) was performed on the FM1 design to assess compliance with mount stability requirements. The analysis considered steady-state thermal conditions with a temperature of $20^{\circ}\mathrm{C}$ applied at the interface with the BlueMUSE main mechanical structure and $26^{\circ}\mathrm{C}$ applied to the mirror surface. The mirror was modeled in Zerodur, the ceramic slopes in EKasic silicon carbide, while the remaining machined components were evaluated using both stainless steel and Invar 36 in order to compare their thermo-mechanical performance.

The results, presented in Figure~\ref{fig:FEA}, predict a total mirror angular variation of $4.8~\mathrm{arcsec}$ when stainless steel is used as the main mount material. Replacing these components with Invar 36 reduces the predicted angular variation to $0.17~\mathrm{arcsec}$ due to its significantly lower coefficient of thermal expansion. The analysis indicates that the Invar 36 design provides sufficient margin to satisfy the updated thermal stability requirement of $0.5~\mathrm{arcsec}$. 

\subsection{Electronics and Software Architecture}
\label{sec:electronics_software}

\begin{figure}[!h]
    \centering
    \includegraphics[width=\textwidth]{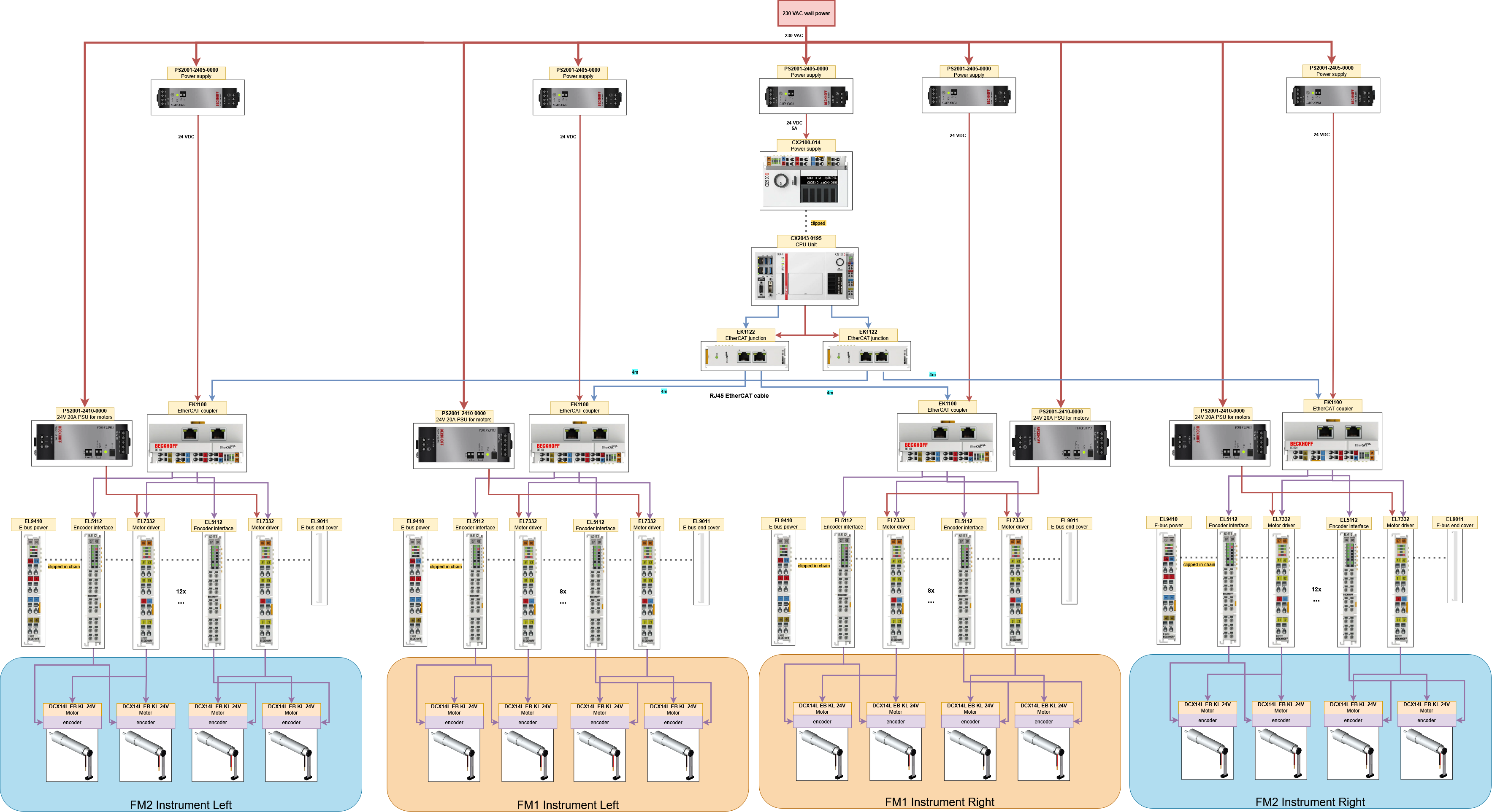}
    \caption{Folding mirrors electronics showing the Beckhoff architecture for all 32 FMs.}  % Single overall caption
    \label{fig:Beckhoff}
\end{figure}

The FM1 mechanism is actuated by two motorized axes, each equipped with a DC motor, an encoder, and limit switches for homing. 

In the BlueMUSE architecture, the FM1 control electronics will be implemented within the Beckhoff EtherCAT environment as depicted in Figure \ref{fig:Beckhoff}. Each DC motor can be driven by an EL7332 DC motor controller terminal, while the motor encoders can be read through EL5112 encoder interface terminals. The Beckhoff PLC would execute the axis-control logic, including homing, calibrated position commands, and communication with the instrument-level control system. The Beckhoff architecture is integrated in the instrument using a 4 branch star distribution, each branch covering a section of FM1s as shown in Figure \ref{fig:Beckhoff}. This allows for more efficient power and data communication, and improved cable management withing the BlueMUSE enclosure.

The folding mirrors are expected to be adjusted in one go, and only occasionally during instrument alignment or recalibration. The control strategy can therefore remain slow before observation. The mechanism is operated open-loop, therefore each commanded movement should be preceded by a homing sequence. This suppresses the accumulation of positioning errors due to backlash and mechanical flexure under load. After homing, the motor moves from the reference position toward the requested target, ensuring that the mechanical transmission is always loaded consistently. Since the mechanical transmission may not be perfectly linear over the full angular range, the final control should rely on a calibration look-up table. 

The software architecture can thus be divided into three layers. The first layer is the low-level motor-control layer, responsible for motor current, velocity, and position control. The second layer is the axis-control layer, which manages homing, limit-switch handling, and conversion from requested mirror angle to motor position. The third layer is the instrument-level interface, which receives alignment corrections from the BlueMUSE control system and sends the corresponding position commands to the FM1 axes.
\section{FM1 Prototype and Test Setup}
\label{sec:prototype_test_setup}

\subsection{Prototype}
\label{sec:prototype}

\begin{figure}[!h]
    \centering
    \includegraphics[width=\textwidth]{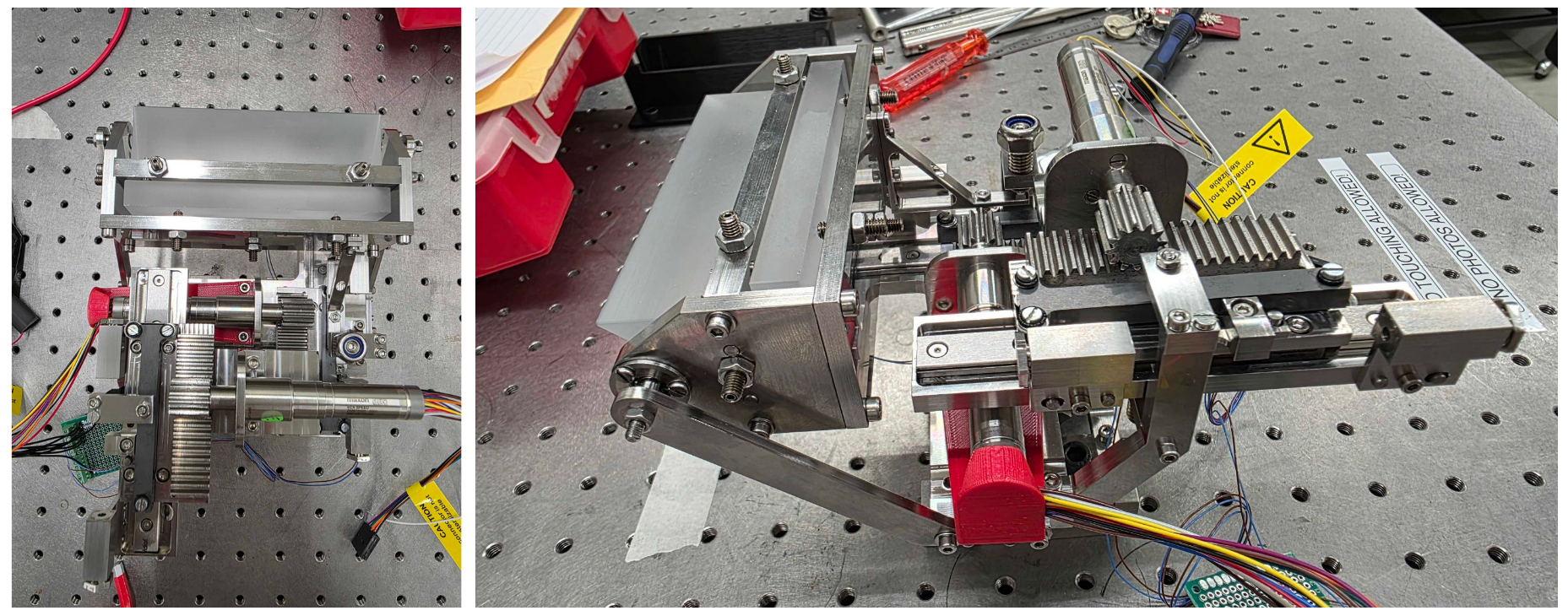}
    \caption{FM1 prototype }  % Single overall caption
    \label{fig:proto}
\end{figure}
A first FM1 prototype was manufactured and assembled within the EPFL Physics department in order to validate the proposed mechanical architecture and characterize its angular repeatability and stability. 

The prototype shown in Figure \ref{fig:proto} was manufactured primarily in stainless steel for the first experimental campaign. This material choice allows rapid prototyping and mechanical validation of the concept. For the final instrument design, Invar 36 is considered as the baseline material for the structural components, due to its low coefficient of thermal expansion and the resulting reduction of thermally induced angular drift, as seen in Figure \ref{fig:FEA}. 

The first prototype was used to identify critical mechanical effects that must be considered in the next design iteration. In particular, backlash in the rack-and-pinion transmission, friction at the ball-slope interface, and slope flexing under the ball bearing load. These effects motivated the use of stiffer and more wear-resistant slope materials, such as EKasic silicon carbide, as well as the addition of improved mechanical adjustability for calibration of the mirror zero position.

\subsection{Optical Characterization Test Setup}
\label{sec:optical_test_setup}

\begin{figure}[!h]
    \centering
    \begin{minipage}[h]{0.47\textwidth}
        \centering
        \includegraphics[width=\textwidth]{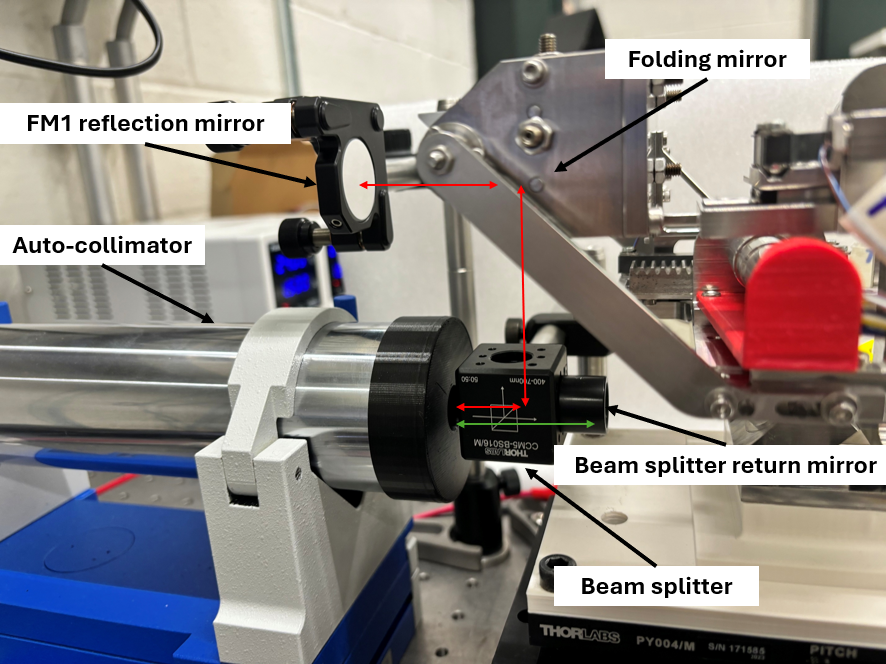}
        \caption{}  % Caption for the first image
        \label{fig:current_testbench}
    \end{minipage}
    \hfill
    \begin{minipage}[h]{0.47\textwidth}
        \centering
        \includegraphics[width=\textwidth]{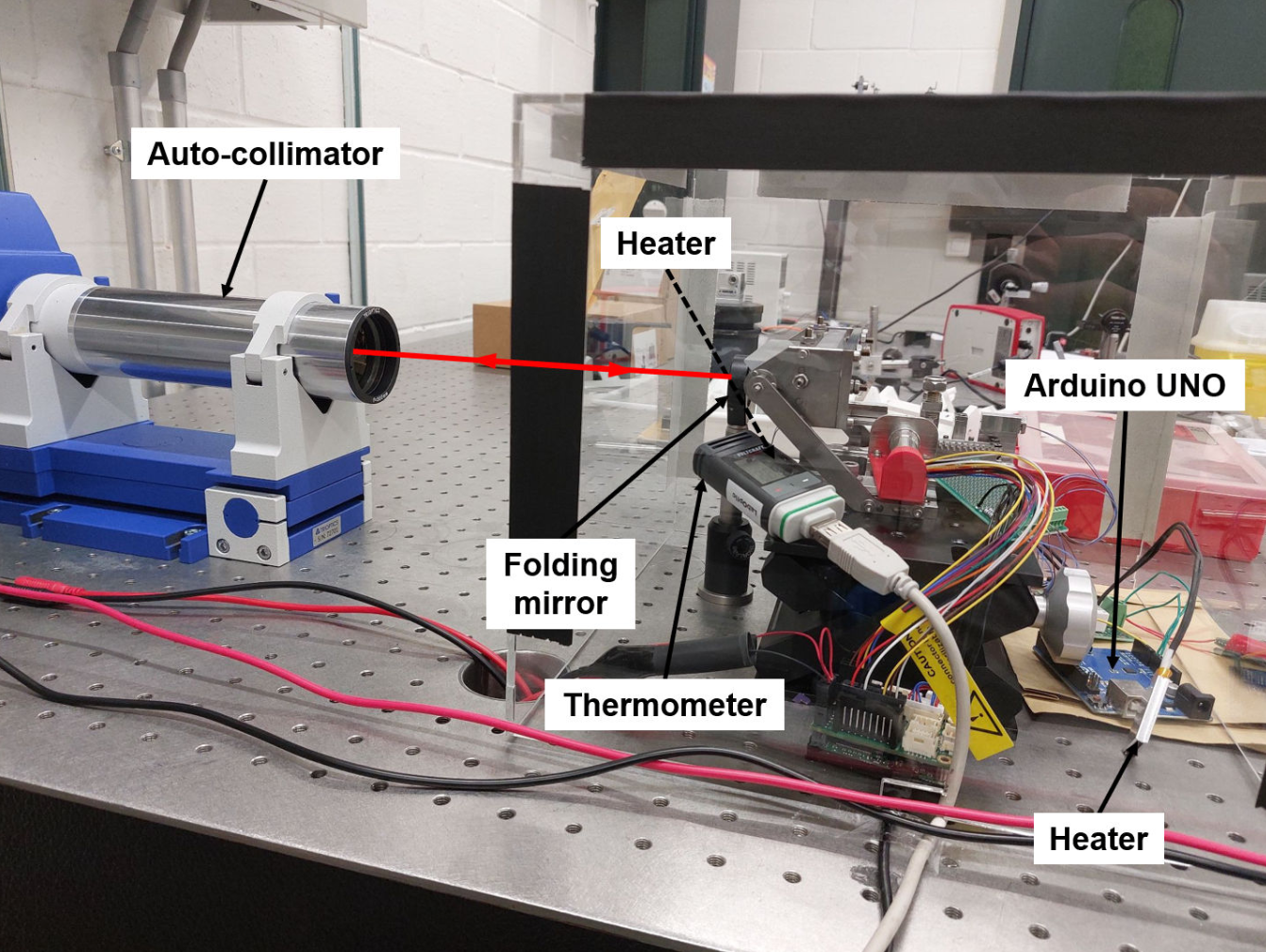}
        \caption{}  % Caption for the second image
        \label{fig:old_testbench}
    \end{minipage}
    \caption*{Folding mirror 1 (FM1) testbenches: (a) FM1 repeatability testbench with differential beam measurements using a beam splitter, reference mirror and autocollimator, the reference beam is drawn in green and the beam reflected by FM1 is drawn in red; (b) FM1 stability testbench with simple beam measurements using autocollimator, heaters, and thermometer.}  % Single overall caption
\end{figure}

The angular performance of the FM1 prototype was measured using an autocollimation-based optical testbench, shown in Figure~\ref{fig:current_testbench}. The purpose of this test bench is to measure the angular displacement of the mirror with sufficient resolution to validate the FM1 repeatability and stability requirements. The autocollimator emits a light beam that is first reflected at 90 degrees by the beam splitter towards FM1, where it is reflected at 90 degrees again. The beam then travels towards a vertically placed fixed mirror (FM1 reflection mirror) where it is reflected back towards the FM1 mirror, the beam splitter, and finally returning to the autocollimator (red beam on Fig. \ref{fig:current_testbench}). Another beam of light goes straight into the beam splitter and is reflected by a small mirror attached to the right end of the beam splitter, providing a reference signal to which the FM1 light path signal is compared to (green beam on Fig. \ref{fig:current_testbench}). The autocollimator then measures the angular deviation between the straight beam reflected by the small reflection mirror, and the beam reflected by FM1. 

This measurement method allows the dissociation of any autocollimator mount instability, which might happen at a micro-level due to the instrument's mechanics, and the light path is such that the angular deviation measurement is independent of small parasitic movements affecting the beam splitter. 

Because the beam is reflected twice on the FM1 mirror in this configuration, the measured angle is not directly equal to the mirror rotation. In the FM1 test configuration, the beam reflects twice on the tested mirror, so the returned beam deviation is proportional to $4\theta$. Therefore, the angle reported by the autocollimator must be divided by four to recover the actual FM1 mirror rotation.

For thermal stability measurements, the FM1 prototype was placed inside a transparent acrylic enclosure (see Fig. \ref{fig:old_testbench}). The enclosure acts as a small thermal chamber while allowing the autocollimator beam to pass through the optical path. The enclosure was manufactured from laser-cut acrylic panels and assembled without glue in order to avoid clouding of the transparent surfaces.

A re-designed testbench is currently being manufactured, including a PEEK plate, which is a highly temperature invariant material, that will be supporting the entirety of the testbench and limiting small high frequency thermal gradients that have been observed between the optical table and the setup, and a larger acrylic enclosure covered in thermal padding, allowing the entire testbench to be enclosed. This will prevent having a material interfering with the light path. Additionally, the optical table will be mounted on pressurised feet to eliminate vibrations linked to environmental factors, as these vibrations create noise in the measured data in the order of $1''$.

A thermal control setup was implemented to impose cyclic temperature variations around the mechanism. The setup used two $6~\mathrm{W}$ resistive heaters controlled by an Arduino UNO through a relay. The temperature inside the enclosure was monitored using a DS18B20 digital temperature sensor with a resolution of approximately $0.06^\circ\mathrm{C}$. Additional temperature logging was performed using a portable thermometer placed inside the enclosure during the tests. This setup allowed the thermal response of the mechanism and the test bench to be evaluated under imposed temperature variations.

\subsection{Measurement Procedures}
\label{sec:measurement_procedures}

Three main experimental procedures were used to characterize the FM1 prototype: repeatability, ambient stability, and thermal stability.
Repeatability and ambient stability measurements were made separately for the 2 motorized axes, while thermal stability measurements were conducted with only the $\Theta X_{FM}$ axis mounted, as this was the only one assembled at the time of testing.

We use a motorized tip tilt stage from Thorlabs to position the test setup in the autocollimator's narrow field of view, and to ajust the closeness of the reference and FM1 beam. 

For repeatability measurements, the mechanism was first referenced using a homing sequence based on mechanical limit switches. The motor was commanded toward the limit switch until contact was detected, then reversed at a lower speed until the switch was released. The release point was used as the reference position. This procedure is advantageous because the reversal during homing removes the effect of backlash before the positioning motion. After homing, the motor was commanded to move by a fixed number of encoder increments, and the resulting mirror angle was measured with the autocollimator. The same sequence was repeated over several cycles to quantify one-shot repeatability.

For ambient stability measurements, the mechanism was left unactuated over several days while the autocollimator continuously recorded the mirror angle. These measurements were used to characterize thermal dependency and long-term drift. The measured signal was processed using a rolling average to estimate the mean angular drift, while the residual signal was used to quantify measurement noise.

For thermal stability measurements, the FM1 prototype was placed inside the acrylic enclosure and subjected to a controlled temperature variation. The mirror angle and enclosure temperature were recorded simultaneously. The objective was to evaluate whether the FM1 mechanism could maintain its angular stability under a representative temperature variation. 

%A separate measurement of the test bench alone was also performed in order to quantify the contribution of the optical setup and support structure to the measured angular drift.

\section{Results and Discussion}
\label{sec:results_discussion}

The FM1 prototype was evaluated with respect to the main angular performance requirements: one-shot repeatability, ambient stability, and thermal stability. The repeatability requirement is associated with the ability of the mechanism to reach a commanded angular position in a single motion after referencing. The stability requirements concern the angular drift of the mirror when the mechanism is not actuated, both under ambient laboratory conditions (under a $\pm 2 ^\circ C$ thermal load) and under large simulated seasonal temperature variations of $\pm 10^\circ C$.

\subsection{FM1 Repeatability}
\label{sec:fm1_repeatability}

\begin{figure}[!h]
    \centering
    \begin{minipage}[h]{0.47\textwidth}
        \centering
        \includegraphics[width=\textwidth]{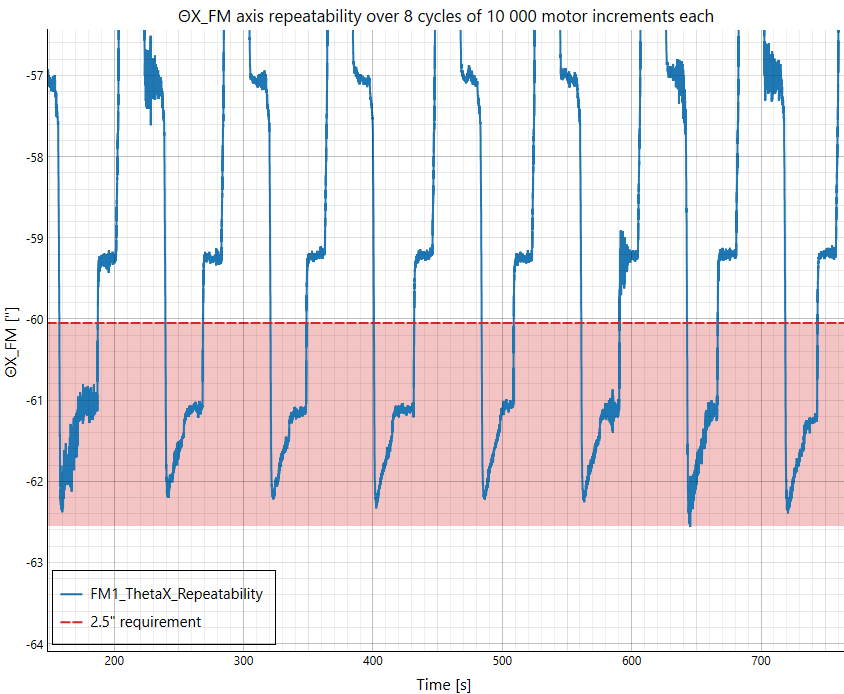}
        \caption*{(a)}  % Caption for the first image
        \label{fig:Theta_X_repeatability}
    \end{minipage}
    \hfill
    \begin{minipage}[h]{0.47\textwidth}
        \centering
        \includegraphics[width=\textwidth]{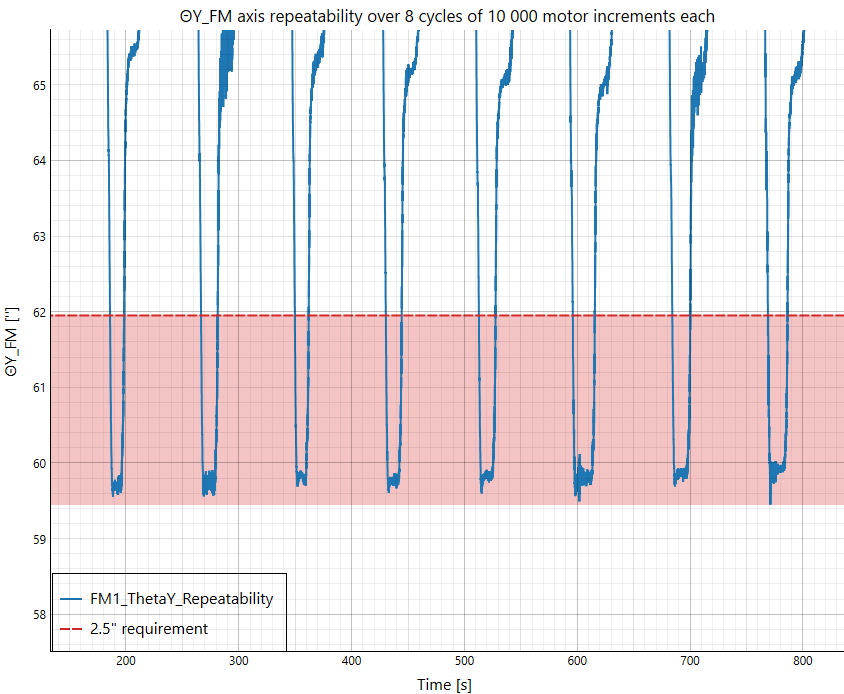}
        \caption*{(b)}  % Caption for the second image
        \label{fig:Theta_Y_repeatability}
    \end{minipage}
    \caption{Repeatability of homing cycles on  the $\Theta X_{FM}$ (a) and on the $\Theta Y_{FM}$ (b) axes. The $2.5''$ repeatability margin is drawn in red. The angular values displayed on the plots have already been divided by four (see section \ref{sec:optical_test_setup}), they thus correspond directly to the FM1 tilt angle.}
    % Single overall caption
    \label{fig:repeatability}
\end{figure}

The repeatability of the FM1 prototype was evaluated using the homing-based positioning strategy described in Section~\ref{sec:measurement_procedures}. After each homing sequence, the motor was commanded to move to the same target position, and the difference between the final mirror angle and the reference beam returned to the small beam splitter mirror was measured with the autocollimator.

The results in Figure \ref{fig:repeatability} showed that the $\Theta X_{FM}$ axis reaches $0.5''$ one-shot repeatability, while the $\Theta Y_{FM}$ axis achieves $0.25''$ variation over several cycles when the motion is performed after homing and in a consistent direction. These values are well below the $2.5''$ repeatability requirement and demonstrates that the mechanical architecture is capable of reaching the required angular precision.

While backlash is present in this mechanism due to the rack and pinion mechanism, and there are elastic effects affecting motion direction changes, these effects have been characterized as repeatable and are suppressed by the zeroing procedure, which will take place before any target maneuver. A calibrated look-up table relating motor encoder position to mirror angle will be implemented to compensate for residual non-linearity in the mechanical transmission. This will also facilitate the FM1s control on the instrument by operators. 

% A slight drift can be observed in the homing sequences of the $\Theta Y_{FM}$ axis, showing that a small error is being amplified with each iteration. This error probably comes from the motor position value. Even though the motor control software returns the same position value after each homing sequence, the autocollimator data says otherwise (Figure \ref{fig:repeatability}). This could be linked to some small errors during the calibration of the motor responsible for the $\Theta Y_{FM}$ axis. The reason the $\Theta X_{FM}$ axis does not exhibit this behavior might be that the motor was installed and calibrated at a later date with a slightly different protocol.
 
% This can be explained by the fact that the currently used homing sequence is a opaque software routine created by the motor's manufacturer, and not necessarily adapted for the stringent repeatability requirements. More testing needs to be done to figure out if tweaking some parameters can eliminate this drift, otherwise, a custom-made motor control software layer will help diagnose this.  

\subsection{FM1 Ambient Stability}
\label{sec:fm1_ambient_stability}

\begin{figure}[!h]
    \centering
    \includegraphics[width=0.55\textwidth]{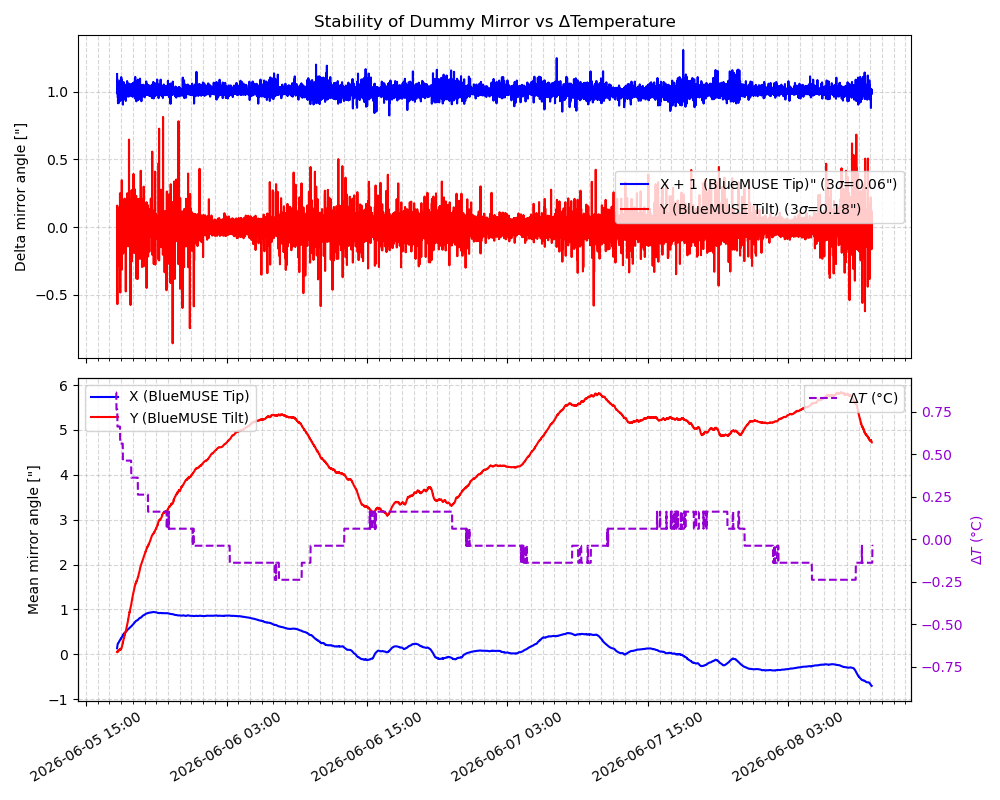}
    \label{fig:FM1_ambient_stability}
    \caption{Folding mirror 1 (FM1) ambient thermal stability evaluated over $\sim$65 hours. The bottom figure shows the mean motion, the top figure shows the noise. "X" and "Y" refer to the autocollimator axes, which correspond to $\theta Y_{FM}$  and $\theta X_{FM}$, respectively.} 
\end{figure}

Ambient stability was measured using the testbench depicted in Figure \ref{fig:current_testbench} by recording the variation between the FM1 mirror angle and the beam splitter's reference beam over long durations without actuating the mechanism. These measurements quantify the angular drift of the mirror under laboratory temperature fluctuations and provide a baseline before applying controlled thermal variations.

The ambient stability measurements showed that, under small temperature variations of $1^\circ C$, the FM1 prototype remained stable within approximately $1.5''$ over several days on the $\Theta X_{FM}$ axis. The $\Theta Y_{FM}$ axis seems to be much more sensitive to temperature variations, its stability margin is approximately $6''$, though this is likely linked to the testbench itself. 

The two regimes in the noise of the $\Theta Y_{FM}$ axis have been identified to be linked with external factors. Indeed, the less noisy intervals roughly correspond to the 00:00 to 05:00 time interval, during which the vibrations from public transports passing close to the lab stop. The testbench noise thus can be evaluated as being $<0.25''$ (the noise amplitude during the "silent" hours).

Dedicated measurements of the optical setup without the FM1 mechanism, and replacing it with a rigid reflection mirror have been conducted on previous iterations of the test bench and showed that it has its own temperature sensitivity. Therefore, ambient stability results include both the FM1 mirror motion and residual motion of the optical setup. The newly developped thermal enclosure and PEEK plate will allow us to eliminate some of these testbenches uncertainties.

\subsection{FM1 Thermal Stability}
\label{sec:fm1_thermal_stability}

Mount stability over thermal cycling was evaluated on our initial testbench shown in Figure \ref{fig:old_testbench} by placing the FM1 prototype inside an acrylic enclosure and applying a cyclic temperature variation using the two heaters. The objective was to reproduce a representative thermal perturbation and measure the resulting angular drift of the mirror.

During the thermal stability test, a temperature variation of approximately $3.5^\circ\mathrm{C}$ produced a maximum measured $\Theta X_{FM}$ motion of approximately $9.5''$, while the measured $\Theta Y_{FM}$ motion remained close to $1''$. The $\Theta X_{FM}$ and $\Theta Y_{FM}$ drifts therefore exceeded the $0.5''$ stability requirement in this preliminary test.

\begin{figure}[!h]
    \centering
    \begin{minipage}[h]{0.47\textwidth}
        \centering
        \includegraphics[width=\textwidth]{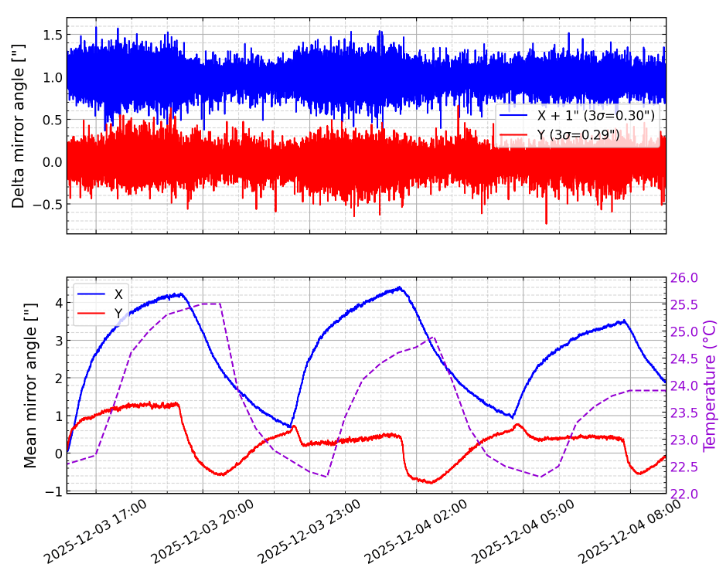}
        \caption*{(a)}  % Caption for the second image
        \label{fig:Testbench_thermal_stab}
    \end{minipage}
    \hfill
    \begin{minipage}[h]{0.47\textwidth}
        \centering
        \includegraphics[width=\textwidth]{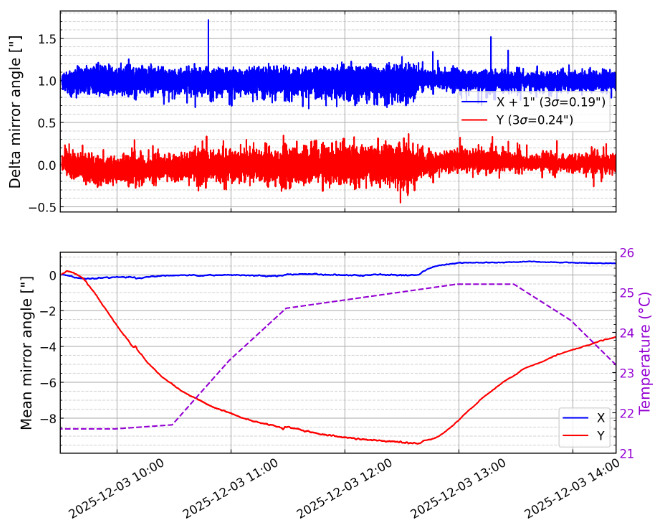}
        \caption*{(b)}  % Caption for the first image
        \label{fig:FM1_thermal_stability}
    \end{minipage}
    \caption{Folding mirror 1 (FM1): (a) Thermal stability of the test bench. The bottom figure shows the mean motion setup, the top figure shows the noise. The X axis shows $\Theta Y_{FM}$ motion of the beam, the Y axis shows $\Theta X_{FM}$ motion; (b) Thermal stability of the FM1 mechanism over 5h. The bottom figure shows the mean motion, the top figure shows the noise. The X axis shows $\Theta Y_{FM}$ motion of the beam, the Y axis shows $\Theta X_{FM}$ motion}  % Single overall caption
    \label{fig:FM1_thermal_stab}
\end{figure}

Several limitations of the test setup justify these results and were identified. First, the optical test bench itself showed angular sensitivity to temperature, as seen in Figure \ref{fig:FM1_thermal_stab}, with drift values an order of magnitude higher than the FM1 stability requirement. Second, the acrylic enclosure was heated by low natural convection, which likely produced temperature gradients around the mechanism. Third, conductive heat transfer through the optical table and support structure may have introduced additional deformation of the test setup. Fourth, the measured beam is affected by the autocollimator's movement and noise, making it difficult to decouple the testbench from the FM1. This was addressed in our current setup with having two return beams and a beam splitter. Finally, the tested prototype used materials and interfaces that differ from the intended final design, including slope materials and adhesive interfaces that may contribute to differential thermal expansion.
\begin{figure}[!h]
    \centering
    \includegraphics[scale=0.4]{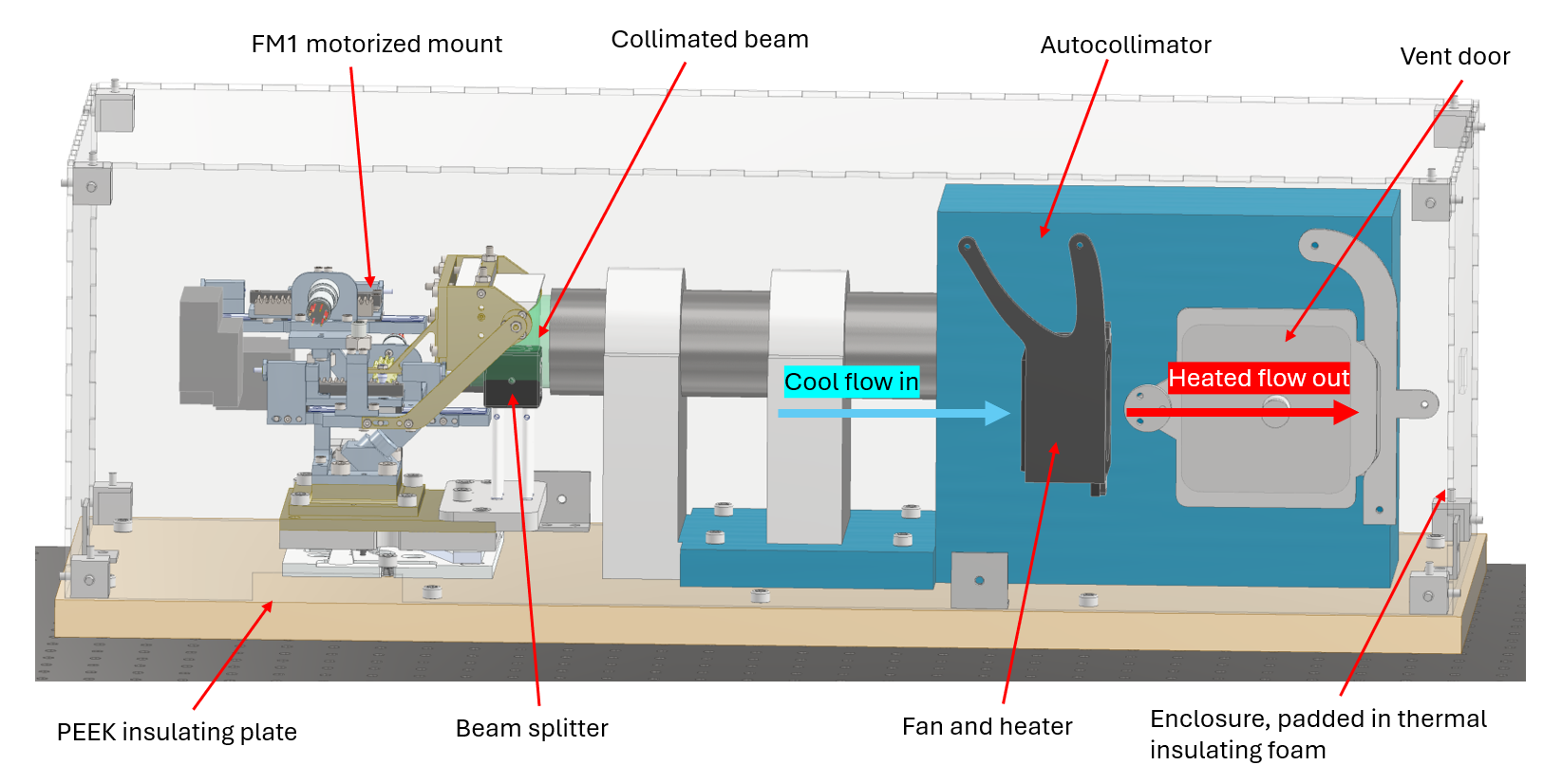}
    \caption{FM1 testbench designed to improve heat distribution and minimize thermal exchanges with the outside environment}  % Single overall caption
    \label{fig:new_testbench}
\end{figure}
A more representative and thermally stable test setup is required before drawing final conclusions. A new testbench has been designed, shown in Figure \ref{fig:new_testbench}, with a PEEK plate supporting the entire testbench to minimize thermal gradients coming from the optical table and with a heater and fan to improve airflow and heat distribution within the enclosure. We are currently implementing 16 temperature sensors on the mirror, mechanism and enclosure to quantify heat distribution and thermal gradients induced by the mount geometry. 
\section{Conclusion}
\label{sec:conclusion}
This paper presented the design, prototyping, and preliminary experimental characterization of a motorized two-degree-of-freedom Folding Mirror 1 mount for the BlueMUSE instrument. The proposed FM1 architecture uses a compact mechanical reduction system based on a rack-and-pinion drive, an inclined ceramic slope, a ball-bearing contact, and a lever arm to convert motor motion into fine angular displacement of the mirror. This approach provides the large motion reduction required to reach arcsecond-level positioning while remaining compatible with the available mechanical envelope of the instrument.

A first stainless-steel prototype was manufactured and tested in order to validate the mechanical concept and identify the main effects influencing angular performance. The repeatability measurements demonstrated that the mechanism is capable of meeting the FM1 one-shot positioning requirement when operated with a consistent homing-based strategy. The measured repeatability reached approximately $0.5''$ for the $\Theta X_{FM}$ axis and $0.25''$ for the $\Theta Y_{FM}$ axis, both below the $2.5''$ requirement. These results confirm that the proposed mechanical architecture can provide the required angular precision, provided that backlash and motion-history effects are controlled through homing and calibrated motion commands.

The stability measurements showed that the prototype is sensitive to thermal effects, but also highlighted the limitations of the initial test setup. Under ambient laboratory conditions, the measured drift remained on the order of a few arcseconds, while controlled thermal tests showed larger angular variations than the $0.5''$ stability requirement. However, measurements without the FM1 mount indicated that a significant part of the observed drift originates from the optical test bench and its thermal sensitivity rather than from the FM1 mechanism alone. The preliminary thermal results therefore cannot yet be used to draw a definitive conclusion on the intrinsic thermal stability of the FM1 design. FEA results show that the design meets the requirements with parts made of invar, and slightly above with parts made of stainless steel. 

These results motivated the development of an improved thermal characterization setup, including a support plate less sensitive to thermal gradients, a larger insulated enclosure, improved air circulation, and additional temperature sensors distributed across the mirror, mechanism, and enclosure. This new setup will allow thermal gradients and testbench-induced angular drift to be better quantified and separated from the actual FM1 response. Future work will also include implementation of the final material choices such as Invar 36 and a mirror in Zerodur, and validation of the calibrated open-loop control strategy within the Beckhoff-based electronics architecture.

Overall, the repeatability performance already satisfies the project requirement, while the remaining work will focus on improving and accurately characterizing the thermal stability of the FM1 mechanism under representative environmental conditions.

\acknowledgments % equivalent to \section*{ACKNOWLEDGMENTS}       
 
The authors would like to thank the Funding LArge international REsearch projects (FLARE) program for supporting the development of the FM1 motorized mount and the associated testbench instrumentation used for its characterization.

The authors also acknowledge the technical staff of the EPFL Physics workshop for their support in the manufacturing of the prototype and experimental setup.

% References
\bibliography{report} % bibliography data in report.bib
\bibliographystyle{spiebib} % makes bibtex use spiebib.bstacv

\end{document}